\documentclass[12pt]{article}
\usepackage{graphicx}
\usepackage{lscape}
\usepackage{citesort}

\begin{document}

\def\ds{\displaystyle}
\def\beq{\begin{equation}}
\def\eeq{\end{equation}}
\def\bea{\begin{eqnarray}}
\def\eea{\end{eqnarray}}
\def\beeq{\begin{eqnarray}}
\def\eeeq{\end{eqnarray}}
\def\ve{\vert}
\def\vel{\left|}
\def\ver{\right|}
\def\nnb{\nonumber}
\def\ga{\left(}
\def\dr{\right)}
\def\aga{\left\{}
\def\adr{\right\}}
\def\lla{\left<}
\def\rra{\right>}
\def\rar{\rightarrow}
\def\nnb{\nonumber}
\def\la{\langle}
\def\ra{\rangle}
\def\ba{\begin{array}}
\def\ea{\end{array}}
\def\tr{\mbox{Tr}}
\def\ssp{{\Sigma^{*+}}}
\def\sso{{\Sigma^{*0}}}
\def\ssm{{\Sigma^{*-}}}
\def\xis0{{\Xi^{*0}}}
\def\xism{{\Xi^{*-}}}
\def\qs{\la \bar s s \ra}
\def\qu{\la \bar u u \ra}
\def\qd{\la \bar d d \ra}
\def\qq{\la \bar q q \ra}
\def\gGgG{\la g^2 G^2 \ra}
\def\q{\gamma_5 \not\!q}
\def\x{\gamma_5 \not\!x}
\def\g5{\gamma_5}
\def\sb{S_Q^{cf}}
\def\sd{S_d^{be}}
\def\su{S_u^{ad}}
\def\ss{S_s^{??}}
\def\sbp{{S}_Q^{'cf}}
\def\sdp{{S}_d^{'be}}
\def\sup{{S}_u^{'ad}}
\def\ssp{{S}_s^{'??}}
\def\sig{\sigma_{\mu \nu} \gamma_5 p^\mu q^\nu}
\def\fo{f_0(\frac{s_0}{M^2})}
\def\ffi{f_1(\frac{s_0}{M^2})}
\def\fii{f_2(\frac{s_0}{M^2})}
\def\O{{\cal O}}
\def\sl{{\Sigma^0 \Lambda}}
\def\es{\!\!\! &=& \!\!\!}
\def\ap{\!\!\! &\approx& \!\!\!}
\def\ar{&+& \!\!\!}
\def\ek{&-& \!\!\!}
\def\kek{\!\!\!&-& \!\!\!}
\def\cp{&\times& \!\!\!}
\def\se{\!\!\! &\simeq& \!\!\!}
\def\eqv{&\equiv& \!\!\!}
\def\kpm{&\pm& \!\!\!}
\def\kmp{&\mp& \!\!\!}


\def\simlt{\stackrel{<}{{}_\sim}}
\def\simgt{\stackrel{>}{{}_\sim}}


\title{
         {\Large
                 {\bf
A comparative study on $B \rar K^\ast \ell^+ \ell^-$ and $B \rar K_0^\ast (1430) \ell^+ \ell^-$ decays  in the
 Supersymmetric Models
                 }
         }
      }

\author{\\
{\small V. Bashiry$^1$\thanks {e-mail: bashiry@ciu.edu.tr}, M.
Bayar$^2$\thanks {e-mail: melahat.bayar@kocaeli.edu.tr}\,\,,  K.
Azizi$^3$\thanks {e-mail: kazizi@dogus.edu.tr}\,\,,} \\ {\small $^1$
Engineering Faculty, Cyprus International University,} \\ {\small
Via Mersin 10, Turkey }\\{\small $^2$ Department of Physics, Kocaeli
University, 41380 Izmit, Turkey}\\{\small $^3$ Physics Division,
Faculty of Arts and Sciences, Do\u gu\c s University,
 }\\{\small Ac{\i}badem-Kad{\i}k\"oy, 34722 Istanbul, Turkey}}

\date{}

\begin{titlepage}
\maketitle
\thispagestyle{empty}

\begin{abstract}
In this paper, we compare the branching ratio and rate difference of
electron channel to muon channel of  $B \rar K_0^\ast (1430) \ell^+
\ell^-$ and  $B \rar K^\ast \ell^+
\ell^-$decays, where $K_0^\ast (1430)$ is the p--wave scalar meson,
in the supersymmetric models. MSSM with $R$ parity is considered since
 considerable deviation from the standard model predictions can be obtained  in $B\rightarrow X_s \ell^-\ell^+$.
Taking $C_{Q1}$ and $C_{Q2}$ about one
which is consistent with the $B\to
K^\ast\mu^+\mu^-$ rate at low dileptonic invariant mass region($1\leq q^2\leq 6$GeV$^2$). It is found
that, firstly,  the $B \rar K_0^\ast (1430) \ell^+ \ell^-$ $(\ell = \mu,\,
\tau)$ decay is measurable at LHC, secondly, in comparison with $B \rar K^\ast \ell^+ \ell^-$
decay a greater deviation in the $B \rar K_0^\ast (1430) \ell^+ \ell^-$ decay can be seen. Measurement of these observables
for the semileptonic rare $B \rar K_0^\ast (1430) \ell^+ \ell^-$, in
particular, at low $q^2$ region can give valuable information about
the nature of interactions within Standard Model or beyond.
\end{abstract}

\end{titlepage}

\section{Introduction}
The Standard Model (SM) is in perfect agreement with all confirmed
collider data, but there is a missing ingredient. The SM is not
regarded as a full theory, since it can not address
 some issues i.e., gauge and fermion mass hierarchy, matter- antimatter
 asymmetry, number of generations, the nature
of the dark matter, the unification of fundamental forces and so on.
For these reasons, the SM
 can be considered as an effective theory of some fundamental theory at low energy.

Supersymmetry (SUSY) is regarded as the most plausible extension of
the SM in order to shed light on some of the issues as mentioned above
\cite{Ellis:2009pz}. It is  an essential ingredient in string theory and
the most-favoured candidate for unifying  all the known interactions
including gravity.  It would help stabilize the hierarchy of mass
scales  between $m_W$ and the Planck mass, by canceling the
quadratic divergences in the radiative corrections to the
mass-squared of the Higgs boson \cite{Witten}.

Two types of study can be conducted to explore supersymmetric
particles (sparticles).  In the direct search, the center of mass
energy of colliding particles has to be increased to produce SUSY
particles at the TeV scale, and hence be accessible to the Large Hadron Collider(LHC).
 On the other hand, we can indirectly investigate SUSY effects. The
sparticles can contribute to the quantum loop. As a result, flavor
changing neutral current (FCNC) transition induced by quantum loop
level can be considered as a good tool for studying the possible
effects of sparticles (there are many studies in this regard, for
the most recent studies see Ref.~\cite{recent} and the references
therein).

The FCNC processes induced by $b \rar s(d)$ transitions are forbidden
 in SM at tree level \cite{R8401,R8402}.  However, they can provide the
most sensitive and stringiest test for the SM at one loop level.
Despite smallness of the branching ratios of FCNC decays, quite
intriguing results have been obtained in ongoing experiments. The
inclusive $B \rar X_s \ell^+ \ell^-$ decay is observed in BaBaR
\cite{R8403} and Belle collaborations. Also these collaborations 
measured exclusive modes $B \rar K \ell^+ \ell^-$
\cite{R8404,R8405,R8406} and $B \rar K^\ast \ell^+ \ell^-$
\cite{R8407}. The experimental results on these decays are in  good
agreement with theoretical estimations \cite{R8408,R8408.1,R8408.2,R8409,R8409.1,R8409.2,R8410,R8410.1,R8410.2} 
 which can be used to constrain new physics (NP) effects.

There is another class of rare decays induced by $b\rar s$
transition, such as $B \rar K_{02}^\ast (1430) \ell^+ \ell^-$ in
which B meson decays into p--wave scalar meson. The decays $B \rar
K_2^\ast (1430) \ell^+ \ell^-$ and $B \rar K_0^\ast (1430) \ell^+
\ell^-$ are studied in \cite{R8411,Aliev:2007rq,Aslam:2009cv}.Transition form factors of these decays in the
framework of light front quark model \cite{R8412} and 3--point QCD
sum rules are estimated in  \cite{R8413}, \cite{R8414}  and
\cite{Aliev:2007rq}, respectively.

In the present work we investigate the possible effects of
sparticles on  the branching ratio of $B \rar K_0^\ast (1430) \ell^+
\ell^-$ decay.

 The paper is
organized as follows: In section 2, we calculate the decay amplitude
of the $B \rar K_0^\ast (1430) \ell^+ \ell^-$ decay within SUSY
models. Section 3 is devoted to the numerical analysis and
discussion of the considered decay and our conclusions.

\section{Decay amplitude of the $B \rar K_0^\ast (1430) \ell^+ \ell^-$ decay in the SUSY models}

The exclusive $B \rar K_0^\ast (1430) \ell^+ \ell^-$ decay is
described at quark level by $b \rar s \ell^+ \ell^-$ transition. The
effective Hamiltonian, that is used to describe the $b \rar s \ell^+ \ell^-$
transition in SUSY models (see, for example,
Ref.~\cite{Aslam:2008hp}), is:

\bea \label{e8401} {\cal H}_{eff} \es {G_F \alpha V_{tb} V_{ts}^\ast
\over 2\sqrt{2} \pi} \Bigg[ C_9^{eff} (m_b) \bar{s}\gamma_\mu
(1-\gamma_5) b \, \bar{\ell} \gamma^\mu \ell + C_{10} (m_b) \bar{s}
\gamma_\mu (1-\gamma_5) b \, \bar{\ell} \gamma^\mu
\gamma_5 \ell \nnb \\
\ek 2 m_b C_7 (m_b) {1\over q^2} \bar{s} i \sigma_{\mu\nu}
q^{\nu}(1+\gamma_5) b \, \bar{\ell} \gamma^\mu \ell + C_{Q_{1}}
\bar{s} (1+\gamma_5)b ~ \bar{\to ell} \ell+ C_{Q_{2}} \bar{s}
(1+\gamma_5)b ~ \bar{\ell}\gamma_5 \ell \Bigg]~,\nnb \\ \eea

SUSY introduces several additional classes of contributions:
 I. gluino, down-type squark loop, II. chargino, up-type
squark loop, III. chargino, up-type squark loop, (Higgs field
attaching to charginos) and IV. neutralino down-type squark
loop\cite{Wang:2003je} accordingly. The neutral Higgs couplings SUSY
contributions are mainly involved via the terms proportional with
$C_{Q_{1,2}}$. These additional terms with respect to the SM   come
from the neutral Higgs bosons (NHBs) exchange diagrams, whose
manifest forms and corresponding Wilson coefficients can be found
in\cite{aslam40,aslam40.1,aslam40.2,aslam41,aslam41.1}. The effects of new scalar and pseudoscalar
type interactions on physical observables come through the terms
which are proportional to the mass of final state leptons. The
effects of the other contributions come through the modification of
known SM Wilson coefficients. The Wilson coefficients $C_7$,
$C_9^{eff}$ and $C_{10}$ are already exist in the SM.
$C_9^{eff}(\hat{s}) = C_9 + Y(\hat{s})$, where $Y(\hat{s}) = Y_{\rm
pert}(\hat{s}) + Y_{\rm LD}$ contains both the perturbative part
$Y_{\rm pert}(\hat{s})$ and long-distance part $Y_{\rm LD}(\hat{s})$
(see Ref.~\cite{R8408,R8408.1,R8408.2}). The explicit expressions of $C_7$,
$C_9^{per}$ and $C_{10}$ in the SM can be found in \cite{R8401}.
$Y_{\rm LD}$ is usually parameterized by using Breit--Wigner ansatz,
\bea Y_{\rm LD} = {3\pi \over \alpha^2} C^{(0)} \sum_{V_i=\psi(1s)
\cdots \psi(6s)} \ae_i \, {\Gamma(V_i \rar \ell^+ \ell^-) m_{V_i}
\over m_{V_i}^2- q^2-im_{V_i}\Gamma_{V_i}} ~, \nnb \eea where
$\alpha$ is the fine structure constant and $C^{(0)}=0.362$.

The phenomenological factors $\ae_i$ for the $B \rar K(K^\ast)
\ell^+ \ell^-$ decay can be determined from the condition that they
should reproduce correct branching ratio relation \bea {\cal B} (B
\rar J/\psi K(K^\ast) \rar K(K^\ast) \ell^+ \ell^-) = {\cal B} (B
\rar J/\psi K(K^\ast)) {\cal B} (J/\psi \rar \ell^+ \ell^-)~, \nnb
\eea  the right--hand side is determined from experiments.
Using the experimental values of the branching ratios for the $B
\rar V_i K(K^*)$ and $V_i \rar \ell^+ \ell^-$ decays, for the lowest
two $J/\psi$ and $\psi^\prime$ resonances, the factor $\ae$ takes
the values: $\ae_1=2.7,~\ae_2=3.51$ (for $K$ meson), and
$\ae_1=1.65,~\ae_2=2.36$ (for $K^\ast$ meson). The values of $\ae_i$
used for higher resonances are usually the average of the values
obtained for the $J/\psi$ and $\psi^\prime$ resonances. In order to
determine the branching ratio for the $B \rar K^\ast_0(1430) \ell^+
\ell^-$ decay with the inclusion of long distance effects, 
the measured branching ratio of $B\rar K^*_0(1430) \psi$ is necessary. However,
the mentioned decay has not been measured yet. Therefore, we assume that
  the values of $\ae_i$ are in the order of one. In accordance, we chose $\ae_1=1$ and
$\ae_2=2$ and performed numerical calculations with these values.

The Wilson coefficients in the framework of the SUSY can be
different from the their SM values. While the SUSY effects on $C_7$,
which  is proportional to the product of the top and bottom Yukawa
coupling constant, $m_t m_b \tan\beta/\sin^2\beta$, is sizable for
large $\tan\beta$. There are no such effects in the calculation of
$C_9$ and $C_{10}$\cite{Wang:2003je}.

 One has to sandwich Eq.~(\ref{e8401}) between initial
meson state $B(p)$ and final  meson state $K_0^\ast (1430)
(p^\prime)$ in order to obtain the amplitude for the $B \rar
K_0^\ast (1430) \ell^+ \ell^-$ decay. Thus, the matrix elements
$\lla K_0^\ast \vel \bar{s}\gamma_\mu (1-\gamma_5) \ver B \rra$ and
$\lla K_0^\ast \vel \bar{s} i \sigma_{\mu\nu} q^\mu (1+\gamma_5)
\ver B \rra$ are needed. These matrix elements are parameterized in
terms of the form factors as follows: \bea \label{e8402} \lla
K_0^\ast (1430) (p^\prime) \vel \bar{s}\gamma_\mu \gamma_5 b \ver
B(p) \rra \es f_+ (q^2) {\cal P}_\mu + f_-(q^2) q_\mu~, \\
\label{e8403} \lla K_0^\ast (1430) (p^\prime) \vel \bar{s}i
\sigma_{\mu\nu} q^\nu \gamma_5 b  \ver B(p) \rra \es {f_T (q^2)
\over m_B + m_{K_0^\ast}} \big[ {\cal P}_\mu q^2 - (m_B^2 -
m_{K_0^\ast}^2 ) q_\mu \big]~, \eea
 where ${\cal P}_\mu =
(p+p^\prime)_\mu$ and $q_\mu = (p-p^\prime)_\mu$. By multiplying
both sides of Eq.~(\ref{e8402}) with $q^\mu$  the
expression in terms of form factors for $\lla K_0^\ast (1430)
(p^\prime) \vel \bar{s}\gamma_5 b \ver B(p) \rra$  can be obtained. \bea
\label{e8404} \lla K_0^\ast (1430) (p^\prime) \vel \bar{s} \gamma_5
b \ver B(p) \rra \es -\frac{1}{m_{b}-m_{s}}[f_+ (q^2) {\cal P}.q +
f_-(q^2) q^{2}]~,\eea

 Using above
Hamiltonian and definitions of form factors, the decay amplitude for
$ B \rar K_0^\ast \ell^+ \ell^-$  can be written as follows:

\bea \label{e8443} {\cal M}(B \rar K_0^\ast \ell^+ \ell^-) \es {G_F
\alpha V_{tb} V_{ts}^\ast \over 2\sqrt{2} \pi} \Bigg[-A_{1} P_{\mu}
\bar{\ell} \gamma^\mu \ell -A_{2} P_{\mu} \bar{\ell} \gamma^\mu
\gamma_5 \ell -A_{3} \bar{\ell} \gamma_{5} \ell -A_{4} \bar{\ell}
\ell \Bigg]~,  \eea where \bea A_{1} &=& C_{9} f_{+} + \frac{2 m_{b}
C_{7}f_{T}}{m_{B}+m_{K_{0}^{*}}} \nnb \\A_{2} &=& C_{10} f_{+} \nnb
\\A_{3} &=& 2 C_{10} m_{\ell} f_{-}+\frac{C_{Q_{2}}}{m_{b}-m_{s}}
[(m_{B}^{2}+m_{K_{0}^{*}}^{2}) f_{+}+q^2 f_{-}]\nnb \\A_{4} &=&
\frac{C_{Q_{1}}}{m_{b}-m_{s}} [(m_{B}^{2}+m_{K_{0}^{*}}^{2})
f_{+}+q^2 f_{-}]. \nnb \eea

Using Eqs. (\ref{e8401})--(\ref{e8443}), we get the following
expression for the differential decay width: \bea \label{e8422} {d
\Gamma  \over dq^2} &=& {G_{F}^2 \alpha^2 \over 8192 m_B \pi^5} \vel
V_{tb} V_{ts}^\ast \ver^2 \upsilon \sqrt{\lambda(1,r,\hat{s})}
\Bigg\{ \frac{4}{3}(\vel A_{1}\ver^2+\vel A_{2}
\ver^2)(-3+\upsilon^2)\Bigg[q^{4}\nnb
\\&-&2q^{2}
 (m_{B}^{2}+m_{K_{0}^{*}}^{2})+(m_{B}^{2}-m_{K_{0}^{*}}^{2})^2
\Bigg]+ 16\vel A_{2} \ver^2
m_{\ell}^{2}\Bigg[q^2-2(m_{B}^{2}+m_{K_{0}^{*}}^{2})\Bigg] \nnb \\
&-&4 q^2 \vel A_{3} \ver^2+4 \vel A_{4} \ver^2 (4m_{\ell}^{2}-q^2)-6
m_{\ell}(A_{2}A_{3}^{*}+A_{2}^{*}A_{3})(m_{B}^{2}-m_{K_{0}^{*}}^{2})
\Bigg\}~,\eea where $\hat{s} = {q^2 \over m_B^2},~ v = \sqrt{1 - {4
m_\ell^2 \over q^2}},~r = {m_{K_0^\ast}^{2}/ m_B^{2}},$ and
$\lambda(1,r,\hat{s}) = 1 + r^2 + \hat{s}^2 - 2 \hat{s} - 2
(1+\hat{s})$.

\section{Numerical results}

In this section, we present the branching ratio  for the both $B \rar
K_0^\ast(1430)$ and  $B \rar K^\ast $ channel for muon and tau leptons. We investigate the
rate difference of electron channel to muon channel. The main input
parameters are the form factors for which we use the results of
three-point QCD sum rules~\cite{Aliev:2007rq}.

The values of the form factors at $q^2=0$ are \cite{Aliev:2007rq}
\bea \label{e8423}
f_+(0) \es \phantom{-} 0.31 \pm 0.08~, \nnb \\
f_-(0) \es -0.31 \pm 0.07~, \nnb \\
f_T(0) \es -0.26 \pm 0.07~, \eea where the errors are due to the
variation of Borel parameters.
%

The best fit for the $q^2$ dependence of the form factors can be
written in the following form: \bea \label{e8425} f_i(\hat{s}) =
{f_i(0)\over 1 - a_i \hat{s} + b_i \hat{s}^2}~, \eea where $i=+$,
$-$ or $T$ and $\hat{s} = q^2/m_B^2$. The values of the parameters
$f_i(0)$, $a_i$ and $b_i$   are specified in Table 1.
\begin{table}[h]
\renewcommand{\arraystretch}{1.5}
\addtolength{\arraycolsep}{3pt}
$$
\begin{array}{|l|ccc|}
\hline
& f_i(0) & a_i & b_i \\ \hline
f_+ &
\phantom{-}0.31 \pm 0.08 & 0.81 & -0.21 \\
f_- &
-0.31\pm 0.07 & 0.80 & -0.36 \\
f_T &
-0.26\pm 0.07 & 0.41 & - 0.32 \\ \hline
\end{array}
$$
\caption{Form factors for $B \rar K_0^\ast(1430) \ell^+ \ell^-$
decay in a three--parameter fit.}
\renewcommand{\arraystretch}{1}
\addtolength{\arraycolsep}{-3pt}
\end{table}

The full kinematical interval of the dilepton invariant mass $q^2$ is $4
m_\ell^2 \le q^2 \le (m_B - m_{K_0^\ast})^2$ for which the long
distance effects (the charmonium resonances) can give substantial
contribution
by the two low lying resonances
$J/\psi$ and $\psi^\prime$, in the interval of $8~GeV^2\le q^2 \le
14~GeV^2$. In order to minimize the hadronic uncertainties we
discard this subinterval by dividing the kinematical region of $q^2$
for muon:
\bea
\begin{array}{cl}
\mbox{\rm I} & 4 m_\ell^2 \le q^2 \le (m_{J\psi} - 0.02~GeV)^2~,\\ \\
\mbox{\rm II} & (m_{J\psi} + 0.02~GeV)^2 \le q^2 \le
(m_{\psi^\prime} - 0.02~GeV)^2~, \\ \\
\mbox{\rm III} & (m_{\psi^\prime} + 0.02~GeV)^2 \le q^2 \le
(m_B-m_{K_0^\ast})^2~. \nnb
\end{array} \nnb
\eea and for tau: \bea
\begin{array}{cl}
\mbox{\rm I} & 4 m_\ell^2 \le q^2 \le (m_{\psi} - 0.02~GeV)^2~,\\ \\
\mbox{\rm II} & (m_{\psi} + 0.02~GeV)^2 \le q^2 \le
(m_B-m_{K_0^\ast})^2. \nnb
\end{array} \nnb
\eea 
The new Wilson coefficients $C_{Q_1}$ and $C_{Q_2}$ are described
in terms of   masses of sparticles i.e., chargino-up-type squark and
NHBs,  $\tan(\beta)$ which is defined as the ratio of the two vacuum
values of the 2 neutral Higgses and $\mu$  which has the dimension
of a mass, corresponding to a mass term mixing the 2 Higgses
doublets. Note that $\mu$ can be positive or negative. Depending on
the magnitude and sign of these parameters, many
options in the parameter space can be considered. However, experimental results i.e., the
rate of $b\rightarrow s \gamma$ and $b\rightarrow s \ell^+ \ell^-$
constrain us to consider the following options:
\begin{itemize}
\item {SUSY I: $\mu$ takes negative value, $C_7$ changes its sign and
contribution of NHBs are neglected.}
\item {SUSY II: $\tan(\beta)$ takes large values while the mass of
superpartners are small i.e., few hundred GeV.}
\item {SUSY III: $\tan(\beta)$ is large and the masses of superpartners are relatively
large, i.e., about 450 GeV or more.}
\end{itemize}

The numerical values of Wilson coefficients  used in our analysis
are referenced from \cite{Aslam:2008hp,aslam12,Bobeth:2007dw}.  In fact,
according to the experimental results obtained by BELLE collaboration\cite{:2009zv}.
  Refs.~\cite{Bobeth:2007dw,Bashiry:2009wh} indicate that for SUSY II in the case of
 muon channel $C_{Q_1}$ and $C_{Q_2}$ should not be greater than $0.5$. In addition to this, in the absence of real experimental constraints on the
FCNC modes in the case of tau channel, we may employ much larger Wilson coefficients
(hence, SUSY effects) than we presented in Tables 2, and 3. Because the Yukawa-driven Higgs coupling implies that $C_{Q}^\tau = m_\tau/
m_\mu C_{Q}^\mu$. The numerical values of Wilson coefficients are collected in Tables 2, and 3.

 In Fig. (1) and (2) we present the
dependence of the differential branching ratio for the $B \rar
K_0^\ast (1430) \ell^+ \ell^-$ and  $B \rar K^\ast  \ell^+\ell^-$ decays, where $\ell=\mu,\, \tau$,  on
$q^2$.
\begin{table}[tbh]
\begin{center}
\begin{tabular}{ccccccc}
\hline Wilson Coefficients & $C_{7}^{eff}$ &  $C_{9}$ &  $C_{10}$ \\
\hline SM & $-0.313$ &  $4.334$ &  $-4.669$\\
\hline SUSY I & $+0.3756$ &  $4.7674$ & $-3.7354$  \\
\hline SUSY II & $+0.3756$ & $4.7674$ &  $-3.7354$
\\ \hline\hline
SUSY III & $-0.3756$ & $4.7674$ & $-3.7354$ \\
\hline\hline
\end{tabular}%
\end{center}
\caption{Wilson Coefficients in SM and different SUSY models but
without NHBs contributions\cite{Aslam:2008hp}.}
\end{table}

\begin{table}[tbh]
\label{NHB}%
\begin{center}
\begin{tabular}{ccccc}
\hline Wilson Coefficients & $C_{Q_{1}}$ & $C_{Q_{2}}$   \\ \hline
SM & $0$  & $0$ \\ \hline SUSY I & $0$ &  $0$  \\
\hline SUSY II & $0.5$\cite{Bobeth:2007dw} $\left( 16.5\right) $\cite{aslam12}& $-0.5$\cite{Bobeth:2007dw} $\left( -16.5\right)
$\cite{aslam12}
\\ \hline\hline
SUSY III & $1.2\left( 4.5\right) $ &  $-1.2\left( -4.5\right) $  \\
\hline\hline
\end{tabular}%
\end{center}
\caption{{}Wilson coefficients corresponding to NHBs contributions
within  SUSY I, II and III models~\cite{Aslam:2008hp}. The values in the bracket are for tau channel.
 Note that the values for SUSY I and III are taken from Ref. \cite{aslam12} and for SUSY II the values taken from \cite{aslam12} and \cite{Bobeth:2007dw}. }
\end{table}

Taking into account the $q^2$ dependence of the form factors given
in Eq. (\ref{e8425}), performing integration over $q^2$, and using
the total lifetime $\tau_B = 1.53 \times 10^{-12}~s$ \cite{R8425},
we get the following results for the branching ratios by considering
short distance contribution:
 \bea {\cal B}(B \rar K_0^\ast(1430) \mu^+ \mu^-) =
\left\{
\begin{array}{lll}
1.05 \times 10^{-7}& \mbox{\rm SUSY I}~,& \\ \\
2.08 \times 10^{-7}& \mbox{\rm SUSY II}~,& \\ \\
1.10 \times 10^{-7}&  \mbox{\rm SUSY III}~,& \end{array} \right.
\nnb \eea

\bea {\cal B}(B \rar K_0^\ast(1430) \tau^+ \tau^-) = \left\{
\begin{array}{lll}
9.54 \times 10^{-10}& \mbox{\rm SUSY I}~,& \\ \\
1.25 \times 10^{-8}& \mbox{\rm SUSY II}~,& \\ \\
2.69 \times 10^{-9}&  \mbox{\rm SUSY III}~.& \end{array} \right.
\nnb \eea

By considering long distance effects in the above--mentioned
kinematical regions, we get the following branching ratios for muon:

\bea {\cal B}(B \rar K_0^\ast(1430) \mu^+ \mu^-) = \left\{
\begin{array}{lll}
1.05 \times 10^{-7}& \mbox{\rm region I}~,& \\ \\
8.98\times 10^{-9}& \mbox{\rm region II}~,& \mbox{\rm for}~\rm SUSY~ I,\\ \\
1.56 \times 10^{-10}&  \mbox{\rm region III}~,& \end{array} \right.
\nnb \eea
\bea {\cal B}(B \rar K_0^\ast(1430) \mu^+ \mu^-) = \left\{
\begin{array}{lll}
1.73 \times 10^{-7}& \mbox{\rm region I}~,& \\ \\
3.71 \times 10^{-8}& \mbox{\rm region II}~,& \mbox{\rm for}~\rm SUSY ~II,\\ \\
3.25 \times 10^{-9}&  \mbox{\rm region III}~,& \end{array} \right.
\nnb \eea and \bea {\cal B}(B \rar K_0^\ast(1430) \mu^+ \mu^-) =
\left\{
\begin{array}{lll}
1.08 \times 10^{-7}& \mbox{\rm region I}~,& \\ \\
1.02 \times 10^{-8}& \mbox{\rm region II}~,& \mbox{\rm for}~\rm SUSY ~III.\\ \\
2.83 \times 10^{-10}&  \mbox{\rm region III}~,& \end{array} \right.
\nnb \eea

and for tau: \bea {\cal B}(B \rar K_0^\ast(1430) \tau^+ \tau^-) =
\left\{
\begin{array}{lll}
5.77 \times 10^{-10}& \mbox{\rm region I}~,& \\ \\
3.43\times 10^{-10}& \mbox{\rm region II}~,& \mbox{\rm for}~\rm
SUSY~ I,
 \end{array} \right. \nnb \eea

 \bea {\cal B}(B \rar K_0^\ast(1430) \tau^+ \tau^-) =
\left\{
\begin{array}{lll}
4.67 \times 10^{-9}& \mbox{\rm region I}~,& \\ \\
5.84\times 10^{-9}& \mbox{\rm region II}~,& \mbox{\rm for}~\rm SUSY~
II,
 \end{array} \right. \nnb \eea
and \bea {\cal B}(B \rar K_0^\ast(1430) \tau^+ \tau^-) = \left\{
\begin{array}{lll}
1.21 \times 10^{-9}& \mbox{\rm region I}~,& \\ \\
1.15\times 10^{-9}& \mbox{\rm region II}~,& \mbox{\rm for}~\rm SUSY~
III.
 \end{array} \right. \nnb \eea
at $f_{K_0^\ast} = 340~MeV$.

\begin{table}
\renewcommand{\arraystretch}{1.5}
\addtolength{\arraycolsep}{3pt}
$$
\begin{array}{|c|c |c||c |c|}
\hline   & B\rightarrow K^* \mu^+ \mu^- &B\rightarrow K_0^\ast \mu^+ \mu^-  \\
\hline
SM~{\mathcal B}(10^{-7}) &1.21^{+0.35}_{-0.39}  & 1.01^{+0.04}_{-0.04}  \\
\hline
\texttt{SUSY I~}{\mathcal B}(10^{-7})  &2.273 &1.05  \\
\hline
\texttt{SUSY II~}{\mathcal B}(10^{-7})  &2.270 & 1.73 \\
\hline
\texttt{SUSY III~}{\mathcal B}(10^{-7})  &0.980 & 1.08  \\
\hline
\texttt{Exp.~}{\mathcal B}(10^{-7}) &1.49^{+0.45}_{-0.40}\pm0.12 \cite{:2009zv}& -  \\
\hline
\end{array}
$$
\caption{Experimentally measured values and integrated values of branching ratio at low dileptonic invariant mass region.}
\renewcommand{\arraystretch}{1}
\addtolength{\arraycolsep}{-3pt}
\end{table}

\begin{table}
\renewcommand{\arraystretch}{1.5}
\addtolength{\arraycolsep}{3pt}
$$
\begin{array}{|c|c |c||c |c|}
\hline   & B\rightarrow K^* \mu^+ \mu^- &B\rightarrow K_0^\ast \mu^+ \mu^- & B\rightarrow K^* \tau^+ \tau^- & B\rightarrow K_0^\ast \tau^+ \tau^- \\
\hline
SM~{\mathcal B}(10^{-7}) &0.158^{+0.004}_{-0.0004}  & 0.015^{+0.002}_{-0.002}&0.11^{+0.01}_{-0.01} & 0.023^{+0.015}_{-0.015} \\
\hline
\texttt{SUSY I~}{\mathcal B}(10^{-7})  &0.181&0.0156 & 0.083& 0.0342 \\
\hline
\texttt{SUSY II~}{\mathcal B}(10^{-7})  &0.184& 0.0325&0.086 & 0.0584 \\
\hline
\texttt{SUSY III~} {\mathcal B}(10^{-7}) &0.173 & 0.0283&0.12& 0.0115 \\
\hline

\end{array}
$$
\caption{Integrated values of branching ratio at high dileptonic invariant mass region($ q^2\geq 14.5$GeV$^2$).}
\renewcommand{\arraystretch}{1}
\addtolength{\arraycolsep}{-3pt}
\end{table}

Our results for low and high $q^2$ regions are shown in the tables 4 and 5.

These results depict that the dominant contribution comes from term
proportional to $C_7$ in region I (low invariant mass region), and
this can be attributed to the existence of the factor $1/q^2$. At
LHCb $10^{11}$--$10^{12}$ pairs are expected  to be produced, the
expected number of events for the $B \rar K_0^\ast(1430) \mu^+
\mu^-$ decay in the low invariant mass region is the order of
$10^4$--$10^5$. Since this region is sensitive to the sign of $C_7$
in the SUSY I model, the study of branching ratio in this region can
provide valuable information about the SUSY effects. In particular,  SUSY I and  SUSY II  can be
distinguished by $B \rar K_0^\ast(1430)$ channel much better than $B \rar K^\ast$ channel(see table 4).  
When value of the branching ratio for the $B \rar
K_0^\ast(1430) \mu^+ \mu^-$ decay is considered both with and 
without long distance effects, valuable results to check structure of the
effective Hamiltonian can be achieved.
The small value of ${\cal B} (B
\rar K_0^\ast(1430) \tau^+ \tau^-)$ can be attributed to the small
phase volume of this decay. Furthermore, SUSY models can enhance the
branching ratio up to one order of magnitude with respect to the SM
values for both $\mu$ and $\tau $ cases. The significant discrepancy
in the non-resonance regions (low $q^2$ and high $q^2$ regions) can
be studied for the effects of not only NHBs but also for NP effects.

Fig.~3 illustrates  the dependency of $R$ in terms of $q^2$ for various
SUSY scenarios for $q^2\geq 4m_{\ell}^2$ region, where $R$ is defined as
follows:\bea R(q^2)=\frac{(d\Gamma/dq^2)(B \rar K_0^\ast(1430) \mu^+ \mu^-)}{(d\Gamma/dq^2)(B \rar K_0^\ast(1430) e^+
e^-)}\eea

Finally, the study of rate difference of muon channel to electron
channel is complimentary work to the studies of other
observables. While SUSY II and SUSY III  approximately coincide
with each other in the study of branching ratio, referred models can be
distinguished by studying the $R$ (see fig.~3). Furthermore, SUSY I
lies in the theoretical error bounds of SM when considering both at
branching ratio (see fig.~1) and $R$ (see fig. 3).

To sum up, we study the semileptonic rare $B \rar K_0^\ast(1430)
\ell^+ \ell^-$ and $B \rar K^\ast\ell^+ \ell^-$ decays in the supersymmetric theories. The results  show that
the branching ratio is very sensitive to the SUSY parameters. The
branching ratio is enhanced up to one order of magnitude with
respect to the corresponding SM values. It is also realized that in the low $q^2$ region the study of
$B \rar K_0^\ast(1430)
\ell^+ \ell^-$ decay is better than $B \rar K^\ast\ell^+ \ell^-$ decay if we try to distinguish SUSY I and SUSY II models.
It is also recognized that while studying the rate difference of electron channel to muon channel, $R$ can be
complimentary to the studies of branching ratio. The results can be
used for indirect search of the SUSY effects in future planned
experiments at LHC.

%

\section*{Acknowledgments}
The authors thank T. M. Aliev for his useful discussions. Special thanks go to Mehmet Toycan for his fruitful contributions to the outline of the paper.

\newpage

\begin{figure}
\vskip 3 cm
    \includegraphics{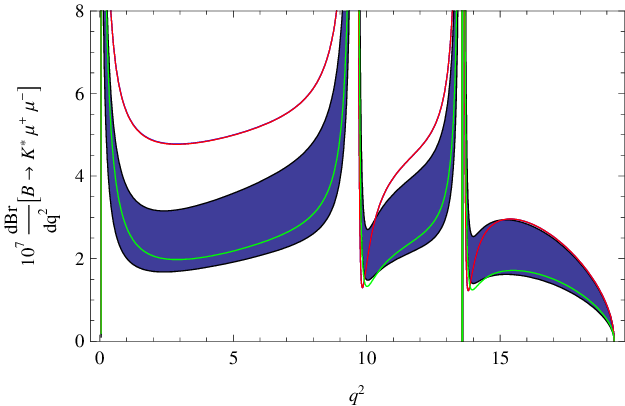}
\includegraphics{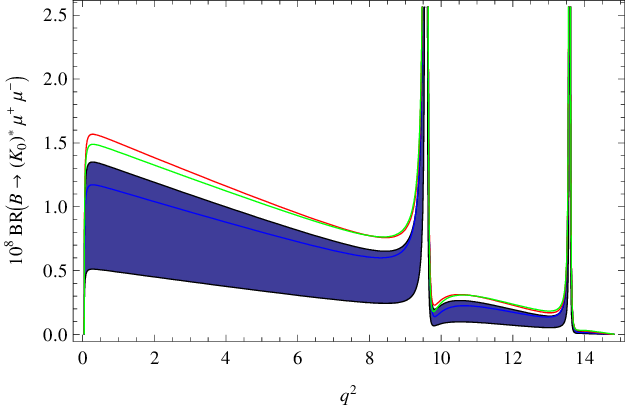}  \vskip 0.5
cm{\small~~~~~~~~~~~~~~~~~~~~~~Fig.
(1a)~~~~~~~~~~~~~~~~~~~~~~~~~~~~~~~~~~~~~~~~~~~~~~~~~~~~~~~~~~~~~~Fig.
(1b)}\caption{Branching ratio of the $B \rar K^\ast \mu^+
\mu^-$ decay and  the $B \rar K_0^\ast(1430) \mu^+
\mu^-$ decay. Black, blue, red and green lines correspond to SM,
SUSY I, SUSY II, SUSY III models, respectively. Blue bound of
the SM is created by the theoretical errors among the formfactors. }
 \end{figure}

\begin{figure}
\vskip 3.5 cm
    \includegraphics{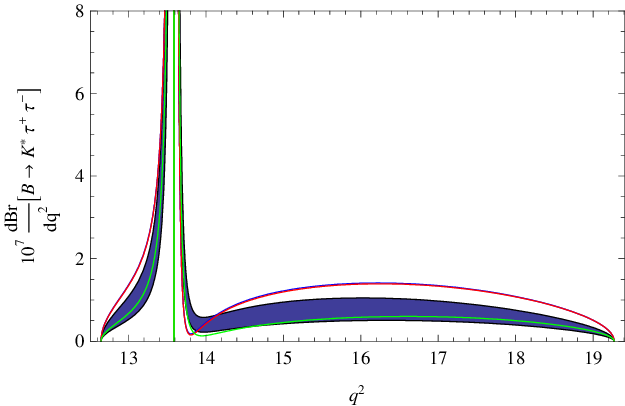}
\includegraphics{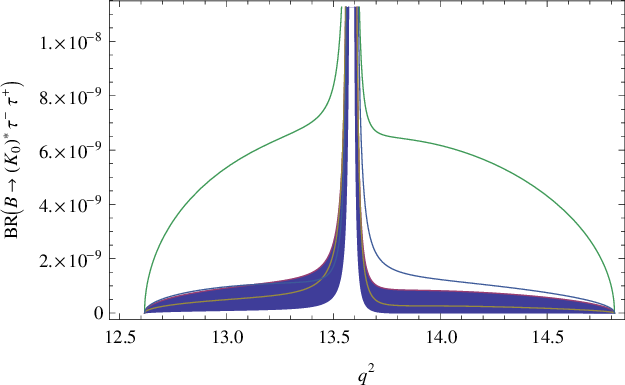}  \vskip 0.5
cm{\small~~~~~~~~~~~~~~~~~~~~~~Fig.
(1a)~~~~~~~~~~~~~~~~~~~~~~~~~~~~~~~~~~~~~~~~~~~~~~~~~~~~~~~~~~~~~~Fig.
(1b)}\caption{The same as Fig. 1 but for tau($\tau$) channel. }
 \end{figure}

\begin{figure}
\vskip 4 cm
    \includegraphics{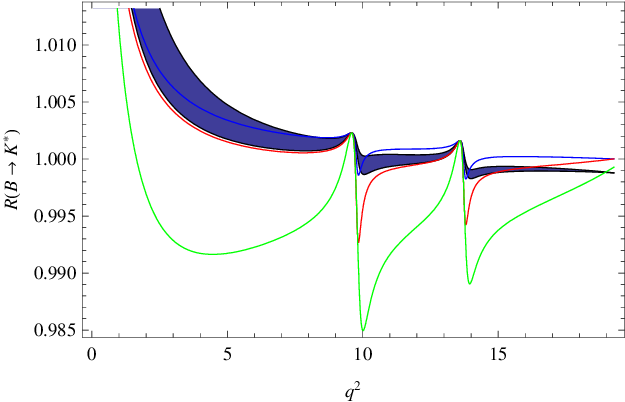}
\includegraphics{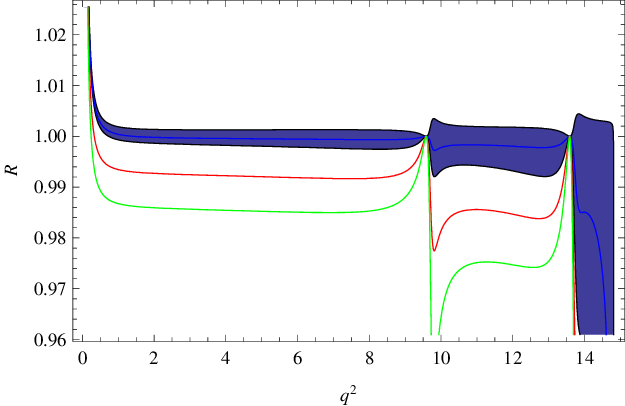}  \vskip 0.5
cm{\small~~~~~~~~~~~~~~~~~~~~~Fig.
(3a)~~~~~~~~~~~~~~~~~~~~~~~~~~~~~~~~~~~~~~~~~~~~~~~~~~~~~~~~~~~~~~~~Fig.
(3b)}\caption{The rate difference of the electron channel to
the muon channel for  the $B \rar K^\ast $ Fig.~(3a)  and the $B \rar K_0^\ast(1430) $ Fig.~(3b) transitions  when $q^2\geq 4m^2_\mu$ region. Blue bound of
the SM is created by the theoretical errors among the formfactors.}
 \end{figure}

\end{document}